\documentstyle[preprint,prb,aps,epsf,eqsecnum]{revtex}
\begin{document}
\tightenlines

\title{Evaluation parameters and Bethe roots for the six vertex 
model at roots of unity}

\author{Klaus Fabricius
\footnote{e-mail Fabricius@theorie.physik.uni-wuppertal.de}}
\address{ Physics Department, University of Wuppertal, 
42097 Wuppertal, Germany}
\author{Barry~M.~McCoy
\footnote{e-mail mccoy@insti.physics.sunysb.edu}}               
\address{ C.N. Yang Institute for Theoretical Physics, 
State University of New York,
 Stony Brook,  NY 11794-3840}
\date{\today}
\preprint{YITPSB-01-42}

\maketitle

\begin{abstract}
We propose an expression for the current form of the lowering operator of the $
{sl}_2$ loop algebra symmetry of the six
vertex model (XXZ spin chain) at roots of unity. 
This operator has poles which correspond to the evaluation parameters
of representation theory which 
are given as the roots of the
Drinfeld polynomial. We explicitly compute these polynomials in
terms of the Bethe roots which characterize the highest weight states
for all values of $S^z$. From these polynomials  
we find that the Bethe roots satisfy sum rules for each
value of $S^z.$

\end{abstract}

\section{Introduction}

Recently it has been demonstrated \cite{paper1}-\cite{paper3} that the 
six vertex model
with periodic boundary conditions
specified by a $2^L\times2^L$ transfer matrix $T(v;\gamma)$ 
\begin{equation}
T_{\{\mu\},\{\mu'\}}(v;\gamma)=\sum_{\lambda_i=\pm 1}
W_{\mu_1,\mu'_1}(\lambda_1,\lambda_2)
W_{\mu_2,\mu'_2}(\lambda_2,\lambda_3)
\cdots W_{\mu_L,\mu'_L}(\lambda_L,\lambda_1)
\label{six}
\end{equation}
with
\begin{eqnarray}
W_{\mu_i,\mu'_i}(\lambda_i,\lambda_{i+1})&&=
{1\over 2}(a+b)\delta_{\mu_i,\mu'_i}\delta_{\lambda_i,\lambda_{i+1}}
+{1\over
2}(a-b)\sigma^z_{\mu_i,\mu'_i}\sigma^z_{\lambda_i,\lambda_{i+1}}\nonumber\\
&&+c(\sigma^+_{\mu_i,\mu'_i}\sigma^-_{\lambda_i,\lambda_{i+1}}
+\sigma^-_{\mu_i,\mu'_i}\sigma^+_{\lambda_i,\lambda_{i+1}}).
\end{eqnarray}
where $\sigma^i$ are the Pauli matrices and 
\begin{equation}
a=i\sinh{1\over 2}(v-i\gamma),~b=-i\sinh{1\over 2}(v+i\gamma),
~c=-i\sinh i\gamma
\end{equation}
has an ${sl}_2$ loop algebra symmetry 
when $\gamma$ satisfies the ``root of unity condition''
\begin{equation}
\gamma=\gamma_0={m\pi\over N} 
\label{root}
\end{equation}
with $m$ and $N$ relatively prime. 
The operator
\begin{equation}
S^z={1\over 2}\sum_{j=1}^L \sigma^z_j
\end{equation}
commutes with the transfer matrix $T(v;\gamma)$
and in  the sector
$S^z\equiv 0 ({\rm mod}~N)$
the ${sl}_2$ loop algebra symmetry was proven in ref. \cite{paper1} 
by demonstrating that the operators with $q=e^{i(\pi-\gamma_0)}$ 
\begin{eqnarray}
S^{\pm(N)}=\sum_{1\leq j_1<\cdots \leq j_N\leq
L}q^{(N/2)\sigma^z}\otimes \cdots \otimes q^{(N/2)\sigma^z}
\otimes \sigma^{\pm}_{j_1}\otimes
q^{((N-2)/2)}\sigma^z\otimes \cdots\nonumber\\
\otimes q^{((N-2)/2)\sigma^z}\otimes
\sigma^{\pm}_{j_2}\otimes q^{((N-4)/2)\sigma^z}\otimes \cdots \otimes
\sigma^{\pm}_{j_N}\otimes q^{-(N/2)\sigma^z}\otimes \cdots \otimes
q^{-(N/2)\sigma^z}
\label{sndf}
\end{eqnarray}
and 
\begin{eqnarray}
T^{\pm(N)}=\sum_{1\leq j_1<\cdots \leq j_N\leq
L}q^{-(N/2)\sigma^z}\otimes \cdots \otimes q^{-(N/2)\sigma^z}
\otimes \sigma^{\pm}_{j_1}\otimes
q^{-((N-2)/2)}\sigma^z\otimes \cdots\nonumber\\
\otimes q^{-((N-2)/2)\sigma^z}\otimes
\sigma^{\pm}_{j_2}\otimes q^{-((N-4)/2)\sigma^z}\otimes \cdots \otimes
\sigma^{\pm}_{j_N}\otimes q^{(N/2)\sigma^z}\otimes \cdots \otimes
q^{(N/2)\sigma^z}
\label{tndf}
\end{eqnarray}
commute (anticommute)with the transfer matrix if $N-m$ is even (odd) 
and satisfy the defining relations for
the Chevalley generators of the $sl_2$ loop algebra. 
In the sectors where 
$S^z \equiv r\neq 0~(\rm {mod}~ N)$ with $1\leq r \leq N-1$
operators such as $(T^+)^r(S^-)^rS^{-(N)}$ commute (anticommute) with
the transfer matrix and
the  ${sl}_2$ loop algebra symmetry was 
inferred from computer computations.

The eigenvalues of the transfer matrix are
polynomials in the variable $e^v.$ In the limit $v\rightarrow 
\pm i\gamma$
the transfer matrix reduces to
\begin{equation}
T(v;\gamma)\rightarrow (\sinh i\gamma)^L \Pi_{\pm}
\left[I-{v\mp i\gamma\over \sinh i\gamma}
\left(H_{\rm XXZ}+{L\cos \gamma\over 2}\right)+O((v\mp i\gamma)^2)\right]
\end{equation}
where $\Pi_{\pm}$ is the left (right) shift operator whose eigenvalues are
$e^{\mp iP}$ where $P$ is the momentum of the  state and
\begin{equation}
H_{\rm XXZ}=-{1\over
2}\sum_{j=1}^L\left(\sigma^x_j\sigma^x_{j+1}+\sigma^y_j\sigma^y_{j+1}+\Delta
\sigma^z_j\sigma^z_{j+1}\right)
\label{ham}
\end{equation}
with
\begin{equation}
\Delta=-\cos \gamma
\end{equation}
is the Hamiltonian of the  XXZ spin chain. Therefore every degeneracy
of the transfer matrix (polynomial in $e^v$) eigenvalues  implies a 
corresponding degeneracy of the (numerical) Hamiltonian eigenvalues.  
However, the converse need not be true and the Hamiltonian (\ref{ham})
is known to have degeneracies such as that caused by invariance under
parity  which the transfer matrix does not have.

The existence  of this symmetry means that the space of eigenstates
states of the transfer matrix can be decomposed into 
finite dimensional representations
of the ${sl}_2$ loop algebra. 
Each finite dimensional representation has a vector $|\Omega>$ with
the property that
\begin{equation}
S^{+(N)}|\Omega>=T^{+(N)}|\Omega>=0 ~~
\end{equation}
and we define $S^z_{\rm max}$ by
\begin{equation}
S^z|\Omega>=S^z_{\rm max}|\Omega>.
\end{equation}
In analogy with the finite Lie algebra $sl_2$ we call this (following
ref. \cite{cp3}) the highest weight vector of the representation.

A fundamental property of all affine Lie algebras \cite{kac}
is that they may be defined either  by a Chevalley basis or by a mode
basis. In the mode basis the elements of the $sl_2$ loop algebra are 
$e(n),~f(n)$ and $h(n),$ where $n$ is an integer, which satisfy the
commutation relations
\begin{eqnarray}
[e(m),f(n)]&&=h(m+n)\nonumber\\
{[}e(m),h(n)]&&=-2e(m+n)\nonumber\\
{[}f(m),h(n)]&&=2f(m+n)
\end{eqnarray}
and there is the
relation to the Chevalley basis of
\begin{eqnarray}
&&e(0)=T^{-(N)},~~e(-1)=S^{-(N)}\nonumber\\
&&f(0)=T^{+(N)},~~f(1)=S^{+(N)}.
\end{eqnarray}
The theory of finite dimensional representations of affine Lie algebras
has been extensively studied in \cite{char}-\cite{cp3} in terms of
this mode basis
where it is shown that
all irreducible finite dimensional representations
are tensor products evaluation representations. 
These evaluation representations are specified by
vectors $|a_j,m_j>$ where for 
all integer $n$ (positive,
negative or zero) 
\begin{eqnarray}
e(n)|a_j,m_j>&&=a_j^ne_{m_j}|a_j,m_j>\nonumber\\
f(n)|a_j,m_j>&&=a_j^nf_{m_j}|a_j,m_j>\nonumber\\
h(n)|a_j,m_j>&&=a_j^nh_{m_j}|a_j,m_j>\nonumber\\
\end{eqnarray}
where $a_j$ are called evaluation parameters and
$e_{m_j},f_{m_j},h_{m_j}$
are a spin $m_j$ representation of $sl_2.$
The evaluation parameters are
shown to be the roots of the Drinfeld polynomial $P_{\Omega}(z)$ defined as
\begin{equation}
P_{\Omega}(z)=\prod_j(z-a_j)^{m_j}
\end{equation}
where the $a_j$ are distinct and
\begin{equation}
P_{\Omega}(z)=\sum_{r\geq 0}\mu_r (-z)^r
\label{drin}
\end{equation}
and for each highest weight state $|\Omega>$ we compute $\mu_r$ from
the eigenvalue equation
\begin{equation}
{T^{+(N)r}\over r!}{S^{-(N)r}\over r!}|\Omega>=\mu_r|\Omega>.
\end{equation}
If the roots of the Drinfeld polynomial (\ref{drin}) are all distinct
only spin 1/2 representations occur. In this case if we call $d_{\Omega}$ the
degree of the Drinfeld polynomial then the number of evaluation parameters
is $d_{\Omega}$ and the number of states with spin $S^z=S^z_{\rm max}-lN$ is
the binomial coefficient
\begin{equation}
{d_{\Omega}\atopwithdelims() l}={\rm number ~of~degenerate~eigenvalues}
\label{binomial}
\end{equation}

In the mode basis we define the  generating function (currents) for
sufficiently large $z$ by
\begin{equation}
E^{-}(z)=\sum_{n=0}^\infty e(n)z^{-n}.
\label{current}
\end{equation}
From the definition of evaluation representation this current will
have poles when $z=a_j.$ In order to show that only spin 1/2
representations occur it is sufficient to demonstrate that there are only
single poles in the current $E(z).$

To compute the Drinfeld polynomial from the definition (\ref{drin})
we need to be in possession of the highest weight vectors.
For a finite Lie algebra these highest weight vectors cannot be
computed from symmetry considerations alone. However for affine Lie
algebras  if the sets of evaluation parameters are all
distinct then   
the algebra will be powerful enough to determine the highest
weight vectors for all multiplets except singlets.

For the six vertex model, however, we may make use of the fact noted in ref.s
\cite{paper1}-\cite{paper3} that the highest weight vectors
of the ${sl}_2$ loop algebra are identical with the
eigenvectors which are
obtained from the solution 
of Bethe's equation for $j=1,\cdots, {L\over 2}-|S^z|$
\begin{equation}
\left({\sinh{1\over 2}(v_j+i\gamma)\over \sinh{1\over 2}(v_j-i\gamma)}
\right)^L=\prod_{l=1\atop l\neq j}^{{L\over 2}-|S^z|}
{\sinh{1\over 2}(v_j-v_l+2i\gamma)\over \sinh{1\over 2}(v_j-v_l-2i\gamma)}
\label{beq}
\end{equation}
In fact there are two distinct ways of specifying the eigenvectors in
terms of the solutions (Bethe roots) of the equation (\ref{beq});
the coordinate Bethe ansatz \cite{bethe}-\cite{liebd} 
which uses explicit forms of the
wavefunctions in coordinate space and the algebraic Bethe ansatz \cite{tf}. 
which produces the wave functions by exhibiting an operator which
creates the wave functions by acting on the state of all spins up. 

In this paper we use the algebraic Bethe
ansatz. In this method of solution of the six vertex model we
introduce the ($ 2\times 2$) monodromy matrix (of operators)
\begin{equation}
M_{\lambda,\lambda'}(v;\gamma)=
\sum_{\lambda_i=\pm 1\atop
i=2,\cdots ,L}
W_{\mu_1,\mu'_1}(\lambda,\lambda_2)
W_{\mu_2,\mu'_2}(\lambda_2,\lambda_3)
\cdots W_{\mu_L,\mu'_L}(\lambda_L,\lambda')
\end{equation}
for which we will use the conventional notation \cite{tf}-\cite{kor}
\begin{equation}
M(v;\gamma)= \left({A(v;\gamma)~~B(v;\gamma)
\atop C(v;\gamma)~~D(v;\gamma)}\right).
\end{equation}
The transfer matrix (\ref{six}) is
\begin{equation}
T(v;\gamma)=A(v;\gamma)+D(v;\gamma).
\label{tac}
\end{equation}
and here and in the rest of the paper we make (for convenience)
the restriction that $L$ be even and thus the eigenvalues of $S^z$ are
integer.

Fundamental to the algebraic Bethe's ansatz are the ``commutation''
relations of the operators ($2^L\times 2^L$ matrices) $A(v;\gamma),~
B(v;\gamma),~C(v;\gamma),~D(v;\gamma)$ of which for our purposes here
we will need the following three
\begin{eqnarray}
[B(v),B(v')]&&=0\label{bb}\\
A(v)B(v')&&=f(v-v')B(v')A(v)+g(v'-v)B(v)A(v')\label{ab}\\
D(v)B(v')&&=f(v'-v)B(v')D(v)+g(v-v')B(v)D(v')\label{db}
\end{eqnarray}
where 
\begin{eqnarray}
f(v)&&=-{\sinh{1\over 2}(v+2i\gamma)\over \sinh(v/2)}\label{f}\\
g(v)&&=-{\sinh i\gamma\over \sinh(v/2)}\label{g}.
\end{eqnarray}
and note the periodicity properties
\begin{equation}
B(v-2\pi i)=-B(v),~~C(v-2\pi i)=-C(v),~~A(v-2\pi i)=A(v),~~D(v-2\pi i)=D(v).
\end{equation}

The eigenvectors of the six vertex model 
for all states where the solutions $v_k$ to Bethe's equation (\ref{beq})
do not lead to undefined factors of $0/0$ are given in this notation by
\begin{equation}
|\{v_k\}>=\prod_{k=1}^nB(v_k;\gamma)|0>
\label{bv}
\end{equation}
where $|0>$ is the unique state with all spins up with
\begin{equation}
S^z|0>={L\over 2}|0>
\end{equation}
and
\begin{equation}
S^z|\{v_k\}>=({L\over 2}-n)|\{v_k\}>.
\label{spin}
\end{equation}

The corresponding eigenvalue of the transfer matrix is
\begin{equation}
t(v)=\sinh^L{1\over 2}(v-i\gamma)\prod_{j=1}^nf(v-v_j)
+\sinh^L{1\over 2}(v+i\gamma)\prod_{j=1}^n f(v_j-v)
\label{teigen}
\end{equation}

If $\gamma$ is a ``generic'' value then the eigenstates (\ref{bv}) are
complete \cite{lanst1}-\cite{lanst2} and there are no factors of $0/0$
in Bethe's equation (\ref{beq}). However, when the root of 
unity condition (\ref{root})
holds it was seen on refs. \cite{paper1}-\cite{paper3} that factors of
$0/0$ do occur in (\ref{beq}) because of solutions which approach the
exact complete N strings first found by Baxter \cite{baxe} where
\begin{equation}
v_k=v_0-2ik\gamma_0~~~{\rm for}~~k=0,1,\cdots,N-1.
\label{string}
\end{equation}

In ref. \cite{paper3} we studied these string solutions of Bethe's
equation (\ref{beq}) in the limit as $\gamma \rightarrow \gamma_0$
and found a set of equations which determines the parameter $v_0$ 
of the exact complete N string (\ref{string}) in this limiting 
sense. However, this is not the only way to proceed because  in the
direct coordinate space solution for the eigenfunctions it was
shown by Baxter in \cite{baxc} that
there are certain equations which vanish
automatically independent of $v_0$ and hence solutions of the string
form (\ref{string}) exist where $v_0$ is arbitrary. It is thus to be
expected that this arbitrary parameter should be closely related to the 
variable $z$ in the current operator (\ref{current})  of the 
affine Lie algebra.

This phenomenon of the automatic vanishing of certain constraint
equations in the coordinate space Bethe's ansatz is mirrored in the
algebraic Bethe's ansatz by the vanishing of the operator \cite{tar} 
\begin{equation}
\prod_{k=0}^{N-1}B(v-2ik\gamma_0)=0
\label{vanish}
\end{equation}
which by (\ref{bv}) should create the exact complete N string (\ref{string}). 

We denote the operator which replaces the vanishing
string creation operator (\ref{vanish}) when the root of unity
condition (\ref{root}) holds by $B^{(N)}(v).$ 
It has the property 
\begin{equation}
T(v)\prod_{j=1}^lB^{(N)}(v'_j)\prod_{k=1}^nB(v_k)|0>=
(-1)^{l(N-m)}\prod_{j=1}^lB^{(N)}(v'_j)T(v)\prod_{k=1}^nB(v_k)|0>
\label{defining}
\end{equation} 
for all $v,v_j'$ and integer $l\geq 1.$
We find (\ref{defining}) is satisfied for $B^{(N)}(v)$ of the form
\begin{eqnarray}
B^{(N)}(v)=&&\sum_{k=0}^{N-1}\left( \prod_{l=0}^{k-1} B(v-2il\gamma_0)
\right)\left( B_{\gamma}(v-2ik\gamma_0)+{X(v-2ik\gamma_0)\over
Y(v)}B_v(v-2ik\gamma_0)\right)\nonumber\\
&&\times\left(\prod_{l=k+1}^{N-1}B(v-2il\gamma_0)\right)
\label{bcurrent}
\end{eqnarray}
where $B_{\gamma}(v)$ and $B_{v}(v)$ specify derivatives of $B(v)$
with respect to $\gamma$ and $v$ respectively.
This operator, with $X(v)$ and $Y(v),$ arbitrary satisfies the commutation
relations
\begin{eqnarray}
[B^{(N)}(v),B(v')]&&=0\label{bnbcomm}\\
{[}B^{(N)}(v),B^{(N)}(v')]&&=0\label{bnbncomm}
\end{eqnarray}
for all values of $v$ and $v'.$

The functions $X(v)$ and $Y(v)$ depend on the highest weight vector on
which the operator acts. Denoting by $v_k$ the solution of
Bethe's equation (\ref{beq}) which specifies the highest weight vector
we find
\begin{eqnarray}
X(v)&&=2i\sum_{l=0}^{N-1}\frac{l \sinh^L{1\over 2}(v-(2l+1)i\gamma_0)}
{\prod_{k=1}^n \sinh{1\over 2}(v-v_k-2il\gamma_0)\sinh{1\over
2}(v-v_k-2i(l+1)\gamma_0)}
\label{xres}
\end{eqnarray}
and 
\begin{equation}
Y(v)=\sum_{l=0}^{N-1}\frac{\sinh^L{1\over 2}(v-(2l+1)i\gamma_0)}
{\prod_{k=1}^n \sinh{1\over 2}(v-v_k-2il\gamma_0)
\sinh{1\over 2}(v-v_k-2i(l+1)\gamma_0)}
\label{yres}
\end{equation}

Because the Chevalley generators $S^{\pm (N)}$ and $T^{\pm (N)}$ (anti) commute
with $T(v)$ the mode operators $e(n)$ and $f(n)$ and their generating
functions
will (anti) commute with $T(v)$ and will therefore obey
(\ref{defining}). We propose that the solution $B^{(N)}(v)$
(\ref{bcurrent}) of (\ref{defining}), (\ref{bnbcomm}) and (\ref{bnbncomm})  
is proportional to the current $E^{-}(z).$
Therefore the zeroes of the function $Y(v)$ 
are the roots of the Drinfeld polynomial 
computed in a fashion totally independent of the definition (\ref{drin}).

We will prove in section two that $B^{(N)}(v)$ satisfies (\ref{defining}).
In section 3 where we will study the polynomial $Y(v)$ and
compute its degree for the various classes of states
specified by 
\begin{equation}
S^z\equiv r~~({\rm mod}~N)
\end{equation}
studied in refs. \cite{paper1}-\cite{paper3}.
We will also see that this polynomial leads to sum rules for each value
of $S^z$ which are not only valid for roots of unity (\ref{root}) but
which hold for all real values of $\gamma.$ 
Examples of these sum rules are
\begin{eqnarray}
S^z&&=0:~~~\sum_{k=1}^{L\over 2}v_k=0,~{\rm or}~\pi i\label{sumrule0}\\
S^z&&=1:~~\{L-(e^{i\gamma}+e^{-i\gamma})\sum_{k=1}^{{L\over2}-1}e^{v_k}\}
            \prod_{k=1}^{{L\over2}-1}e^{v_k}
          =\{L-(e^{i\gamma}+e^{-i\gamma})\sum_{k=1}^{{L\over 2}-1}e^{-v_k}\}
            \prod_{k=1}^{{L\over2}-1}e^{-v_k}\label{sumrule1}\\
S^z&&=2:~~
\left(L(L-1)-2L(e^{i\gamma}+e^{-i\gamma})\sum_{k=1}^{{L\over 2}-2}e^{
v_k}\right.\nonumber\\
&&\left.+(e^{2i\gamma}+e^{-2i\gamma})\sum_{k=1}^{{L\over 2}-2} e^{2v_k}
+(e^{i\gamma}+e^{-i\gamma})^2(\sum_{k=1}^{{L\over 2}-2}e^{v_k})^2\right)
\prod_{k=1}^{{L\over 2}-2} 
e^{v_k}\nonumber\\
&&=\left(L(L-1)-2L(e^{i\gamma}+e^{-i\gamma})\sum_{k=1}^{{L\over 2}-2}
e^{-v_k}\right.\nonumber\\
&&\left.+(e^{2i\gamma}+e^{-2i\gamma})\sum_{k=1}^{{L\over 2}-2} e^{-2v_k}
+(e^{i\gamma}+e^{-i\gamma})^2(\sum_{k=1}^{{L\over 2}-2}e^{-v_k})^2\right)
\prod_{k=1}^{{L\over 2}-2} e^{- v_k}.
\label{sumrule2}
\end{eqnarray}
The sum rule (\ref{sumrule0}) is the XXZ version of a sum rule in the
XYZ model \cite{sumbax} and has recently been studied
numerically \cite{sumnep1}, \cite{sumnep2}. The sum rules for $S^z\neq 0$
have not previously been seen.

In section 4 we 
discuss the relation between these algebraic Bethe's
ansatz computations and the theory of finite dimensional representations
of affine Lie algebras and quantum groups at roots of unity.
Concluding remarks are in sec. 5. 

\section{Derivation of the Operator $B^{(N)}(v)$}

We begin the study of the operator $B^{(N)}(v)$ given by 
(\ref{bcurrent}) by 
proving the commutation relations (\ref{bnbcomm}) and 
(\ref{bnbncomm}) which do not depend on the form of $X(v)$ and $Y(v)$.
We will then use the defining property (\ref{defining}) to determine
the functions $X(v)$ and $Y(v)$ given by (\ref{xres}) and
(\ref{yres}).

\subsection{Proof of (\ref{bnbcomm})}

Write the operator $B^{(N)}(v)$ of (\ref{bcurrent}) as
\begin{equation}
B^{(N)}(v)=B_1^{(N)}(v)+B_2^{(N)}(v)/Y(v)
\label{bdecomm}
\end{equation}
with 
\begin{eqnarray}
B_1^{(N)}(v)&&=\sum_k^N\left( \prod_{l=0}^{k-1} B(v-2il\gamma_0)
\right)\left( B_{\gamma}(v-2ik\gamma_0)\right)
\left(\prod_{l=k+1}^{N-1}B(v-2il\gamma_0)\right)\\
B_2^{(N)}(v)&&=\sum_k^N\left( \prod_{l=0}^{k-1} B(v-2il\gamma_0)
\right)\left(X(v-2ik\gamma_0)B_v(v-2ik\gamma_0)\right)
\left(\prod_{l=k+1}^{N-1}B(v-2il\gamma_0)\right).
\end{eqnarray}
Then by differentiating (\ref{bb}) with respect to $v$ we 
find
\begin{equation}
[B_v(v),B(v')]=0
\label{bvbcomm}
\end{equation}
and hence
\begin{equation}
[B_2^{(N)}(v),B(v')]=0.
\label{b2bcomm}
\end{equation}

It also follows from (\ref{bb}) that
\begin{equation}
[\prod_{l=0}^{N-1}B(u_l),B(v')]=0.
\end{equation}
Differentiating this with respect to $\gamma$ and then setting
$u_l=v-2il\gamma_0$ we find
\begin{equation}
[B_1^{(N)}(v),B(v')]+[\prod_{l=0}^{N-1}B(v-2il\gamma_0),B_{\gamma}(v')]=0.
\end{equation}
Both terms in the second commutator are zero because of the vanishing
condition (\ref{vanish}) and thus
\begin{equation}
[B_1^{(N)}(v),B(v')]=0.
\label{b1bcomm}
\end{equation}
Combining (\ref{b2bcomm}) and (\ref{b1bcomm}) we obtain the desired
result (\ref{bnbcomm}). 
\subsection{Proof of (\ref{bnbncomm})}

Using the decomposition (\ref{bdecomm}) we see that the desired
commutation relation (\ref{bnbncomm}) will follow from the three
separate commutation relations
\begin{equation}
[B_j^{(N)}(v),B_{j'}^{(N)}(v')]=0~~~{\rm for}~~j,j'=1,2
\label{help1}
\end{equation}
 
To prove (\ref{help1}) for  $j=j'=1$ 
we note that from (\ref{bb}) it follows that
\begin{equation}
[\prod_{l=0}^{N-1}B(u_l),\prod_{l=0}^{N-1}B(w_l)]=0
\label{help2}
\end{equation}
and thus
\begin{equation}
{d^2\over d\gamma^2}[\prod_{l=0}^{N-1}B(u_l),\prod_{l=0}^{N-1}B(w_l)]=0
\end{equation}
Then if we first carry out the differentiations, then set
$u_l=v-2il\gamma_0,~w_l=w-2il\gamma_0$ and then use the vanishing condition
(\ref{vanish}) we see that the only non-vanishing terms are those
where one derivative acts on a $B(u_l)$ and one acts on a $B(w_l)$
and thus it follows that
\begin{equation}
[B_1^{(N)}(v),B_1^{(N)}(w)]=0.
\end{equation}

To prove (\ref{help1}) for $j=1,~j'=2$ we differentiate (\ref{help2})
with respect to $\gamma$ and $w_l$  to find
\begin{eqnarray}
&&{d^2\over d\gamma
dw_k}[\prod_{l=0}^{N-1}B(u_l),\prod_{l=0}^{N-1}B(w_l)]\nonumber\\
&&={d\over d\gamma}[\prod_{l=0}^{N-1}B(u_l),
\left(\prod_{l=0}^{k-1}B(w_l)\right)B_w(w_k)\left(\prod_{l=k+1}^{N-1}
B(w_l)\right)=0
\end{eqnarray}
Then as before carry out the differentiation with respect to $\gamma.$
set $v_l=v-2i\l\gamma_0,~w_l=w-2il\gamma_0,$ and use the vanishing
condition (\ref{vanish}) to obtain
\begin{equation}
[B_1^{(N)}(v),\left(\prod_{l=0}^{k-1}B(w_l)\right)B_w(w_k)
\left(\prod_{l=k+1}^{N-1}B(w_l)\right)]=0
\end{equation}
from which (\ref{help1}) follows for $j=1,~j'=2.$

Finally we note that (\ref{help1}) for $j=j'=2$ follows immediately
from (\ref{bb}). Thus we have proven the desired commutation relation
 (\ref{bnbncomm}).

\subsection{Derivation of $X(v)$ and $Y(v)$}  

The functions $X(v)$ and $Y(v)$ are uniquely determined by the
defining property of the operator $B^{(N)}(v)$
(\ref{defining}) with $l=1.$ We begin the derivation by proving the
following four relations:

\vspace{.1in}

{\bf Proposition}

\vspace{.1in}

\begin{eqnarray}
(1)~~~&&A(v)B_v(v')\prod_{l=1}^{N-1}B(v'-2il\gamma_0)
=(-1)^{N+m}B_v(v')\left(\prod_{l=0}^{N-1}B(v'-2il\gamma_0)\right)A(v)\nonumber\\
&&+{(-1)^{N+m}\over 2\sinh{1\over 2}(v'-v)}B(v)
\left(\prod_{l=1\atop \neq 0}^{N-1}B(v'-2il\gamma_0)\right)
A(v')\nonumber\\
&&-{(-1)^{N+m}\over 2\sinh{1\over 2}(v'-v-2i\gamma_0)}B(v)\left(\prod_{l=0\atop
\neq 1}^{N-1}B(v'-2il\gamma_0)\right) A(v'-2i\gamma_0)
\label{rel1}
\end{eqnarray}
\begin{eqnarray}
(2)~~~&&D(v)B_v(v')\prod_{l=1}^{N-1}B(v'-2il\gamma_0)
=(-1)^{N+m}B_v(v')\left(\prod_{l=0}^{N-1}B(v'-2il\gamma_0)\right)D(v)
\nonumber\\
&&+{(-1)^{N+m}\over 2\sinh{1\over 2}(v'-v)}B(v)\left(\prod_{l=0\atop
\neq 0}^{N-1}B(v'-2il\gamma_0)\right) D(v')\nonumber\\
&&-{(-1)^{N+m}\over 2\sinh{1\over 2}(v'-v-2(N-1)i\gamma_0)}B(v)
\left(\prod_{l=1\atop \neq N-1}^{N-1}B(v'-2il\gamma_0)\right)
D(v'-2(N-1)i\gamma_0)
\label{rel2}
\end{eqnarray}
\begin{eqnarray}
(3)~~~&&A(v)B_1^{(N)}(v')=(-1)^{N+m}B_1^{(N)}(v')A(v)\nonumber\\
&&+i\sum_{k=0}^{N-1}{(-1)^{N+m}\over 2\sinh{1\over 2}(v'-v-2ik\gamma_0)}B(v)
\left(\prod_{l=0\atop \neq k}^{N-1}B(v'-2il\gamma_0)\right)
A(v'-2ik\gamma_0)
\label{rel3}
\end{eqnarray}
and
\begin{eqnarray}
(4)~~~&&D(v)B_1^{(N)}(v')=(-1)^{N+m}B_1^{(N)}(v')D(v)\nonumber\\
&&-i\sum_{k=0}^{N-1}{(-1)^{N+m}\over 2\sinh{1\over 2}(v'-v-2ik\gamma_0)}B(v)
\left(\prod_{l=0\atop \neq k}^{N-1}B(v'-2il\gamma_0)\right)
D(v'-2ik\gamma_0)
\label{rel4}
\end{eqnarray}

\vspace{.1in}

{\bf Proof}

\vspace{.1in}

We begin by iterating the basic relations of the algebraic Bethe's
ansatz
(\ref{bb})-(\ref{db}) to obtain
\begin{equation}
A(v)\prod_{l=1}^nB(u_l)=
\Lambda\left(\prod_{l=1}^nB(u_l)\right)A(v)
+B(v)\sum_{k=1}^n\Lambda_k\left(\prod_{l=1\atop \neq
k}^nB(u_l)\right) A(u_k)
\label{abit}
\end{equation}
and
\begin{equation}
D(v)\prod_{l=1}^nB(v_l)={\tilde\Lambda}
\left(\prod_{l=1}^nB(u_l)\right)D(v)
+B(v)\sum_{k=1}^n{\tilde\Lambda}_k\left(\prod_{l=1\atop \neq
k}^nB(u_l)\right) D(u_k)
\label{dbit}
\end{equation}
where
\begin{eqnarray}
&&\Lambda=\prod_{l=1}^n f(v-u_l)\label{lamb},
~~~\Lambda_k=g(u_k-v)\prod_{l=1\atop \neq
k}^nf(u_k-u_l)\label{lambk}\\
&&{\tilde \Lambda}=\prod_{l=1}^n f(u_l-v),~~~
{\tilde \Lambda}_k=g(v-u_k)\prod_{l=1\atop \neq
k}^nf(u_l-u_k)\label{tlambk}
\end{eqnarray}

To prove (\ref{rel1}) we start with (\ref{abit}) 
and differentiate with respect
to $u_1$ to obtain
\begin{eqnarray}
&&A(v)B_v(u_1)\prod_{l=2}^nB(u_l)\nonumber\\
&&=\Lambda B_v(u_1)\left(\prod_{l=2}^n B(u_l)\right) A(v)
+B(v)\Lambda_1\left(\prod_{l=2}^n B(u_l)\right) A_v(u_1)\nonumber\\
&&+B(v)\sum_{k=2}^n\Lambda_kB_v(u_1)
\left(\prod_{l=2\atop l\neq k}^{n}B(u_l)\right)  A(u_k)\nonumber\\
&&+{\partial \Lambda\over \partial v_1}(\left( \prod_{l=1}^n
B(u_l)\right) A(v)+
B(v)\sum_{k=1}^n {\partial \Lambda_k\over \partial v_1}
\left(\prod_{l=1\atop \neq k}^nB(u_l)\right)
A(u_k)
\label{diff1}
\end{eqnarray}
where
\begin{eqnarray}
&&{\partial \Lambda\over \partial u_1}=-f_v(v-u_1)\prod_{l=2}^n
f(v-u_l)\\
&&{\partial \Lambda_1\over \partial
u_1}=g_v(v_1-v)\prod_{l=2}^nf(u_1-u_l)
+g(u_1-v)\sum_{k=2}^nf_v(u_1-u_k)\prod_{l=2 \atop \neq k}^n f(u_1-u_l)\\
&&{\partial \Lambda_k\over \partial u_1}=-g(u_k-v)f_v(u_k-u_1)\prod_{l=2\atop
\neq k}^n f(v_k-v_l)~~{\rm for}~k\neq 1.
\end{eqnarray}
Now after differentiating set $n=N$ and $u_l=v'-2i(l-1)\gamma_0.$ 
Then
\begin{eqnarray}
f(-2i\gamma_0)&&=f(2(N-1)i\gamma_0)=0\\
f_v(-2i\gamma_0)&&=f_v(2(N-1)i\gamma_0)={1\over 2\sinh i\gamma_0}
\end{eqnarray}
and therefore
\begin{eqnarray}
\Lambda=&&\prod_{l=1}^{N}f(v-u_l)
=\prod_{l=1}^N-{\sinh{1\over 2}(v-v'+2il\gamma_0)
\over \sinh{1\over 2}(v-v'+2i(l-1)\gamma_0)}
=(-1)^N{\sinh{1\over 2}(v-v'+2iN\gamma_0)
\over \sinh{1\over 2}(v-v')}\nonumber\\
&&=(-1)^{N+m}\label{lam}\\
\Lambda_k=&&0~~{\rm for}~~k=1,\cdots,N\label{lamk}\\
{\partial \Lambda_1\over \partial u_1}=
&&{(-1)^{N+m}\over 2\sinh{1\over 2}(v'-v)}\\
{\partial \Lambda_2\over \partial u_1}=&&{(-1)^{N+m+1}\over
2\sinh{1\over 2}(v'-v-2i\gamma_0)}\\
{\partial \Lambda_k\over \partial u_1}=&&0~~{\rm for}~~k=3,\cdots.N
\end{eqnarray}
and using the vanishing condition (\ref{vanish})
we find that the only terms in (\ref{diff1}) which do not vanish are the
first term and the last term with $k=1,2.$ Thus we find that
(\ref{diff1}) reduces to (\ref{rel1}) as desired.

To prove (\ref{rel3}) we similarly  differentiate (\ref{abit}) 
with respect to $\gamma$ and then set $n=N$ and $u_l=v'-2(l-1)i\gamma_0.$
the derivative of the left hand
 side of (\ref{abit}) is
\begin{equation}
A(v)\sum_{k=1}^N\left(\prod_{l=0}^{k-1} B(v'-2i\gamma_0)\right)
B_{\gamma}(v'-2ik\gamma_0)
\left(\prod_{l=k+1}^{N-1} B(v'-2il\gamma_0)\right)=A(v)B_1^{(N)}(v')
\end{equation}
where the term with $A_{\gamma}(v_0)$ vanishes because of the
condition (\ref{vanish}). To differentiate the right hand side of
(\ref{abit}) we use both the vanishing condition (\ref{vanish}) and
conditions
(\ref{lam}) and (\ref{lamk}) on $\Lambda$ and $\Lambda_k.$ 
Thus we find that the
derivative of (\ref{abit}) with respect to $\gamma$ reduces to
\begin{equation}
A(v)B_1^{(N)}(v')=(-1)^{N+m}B_1^{(N)}(v')A(v)
+B(v)\sum_{k=1}^N {\partial \Lambda_k\over \partial \gamma}
\left( \prod_{l=1\atop \neq k}B(v'-2il\gamma_0)\right)A(v'-2ik\gamma_0).  
\label{hlprel}
\end{equation}
By differentiating (\ref{lambk}) with respect to $\gamma$ and noting
that 
\begin{equation}
f_{\gamma}(v)=-i{\cosh{1\over 2}(v+2i\gamma)\over \sinh(v/2)}
\end{equation}
we find
\begin{equation}
{\partial \Lambda_k\over\partial \gamma}=i{(-1)^{N+m+1}\over
\sinh{1\over 2}(v'-v-2i(k-1)\gamma_0)}
\end{equation}
we find that (\ref{hlprel}) reduces to (\ref{rel3}) as desired.

The proof of (\ref{rel2}) and (\ref{rel4}) follows in an similar
manner by differentiating (\ref{dbit})

\vspace{.1in}

{\bf QED}

\vspace{.1in}

We now use (\ref{rel1})-(\ref{rel4}), the definition (\ref{bdecomm}) 
of $B^{(N)}(v)$ in
terms of $B_1^{(N)}(v)$ and $B_2^{(N)}(v)$ to find
\begin{eqnarray}
A(v)B^{(N)}(v')=&&(-1)^{N+m}B^{(N)}(v')A(v)\nonumber\\ 
&&-B(v)\sum_{k=0}^{N-1}
\left({X(v'-2i(k-1)\gamma_0)-X(v'-2ik\gamma_0)\over
Y(v')}-2i\right)\nonumber\\
&&\times\left(\prod_{l=1\atop \neq k}^N 
B(v'-2il\gamma_0)\right){(-1)^{N+m} A(v'-2ik\gamma_0)\over 2\sinh{1\over
2}(v'-v-2ik\gamma_0)}\label{amaster} \\
D(v)B^{(N)}(v')=&&(-1)^{N+m}B^{(N)}(v')D(v)\nonumber\\ 
&&-B(v)\sum_{k=0}^{N-1}\left({X(v'-2i(k+1)\gamma_0)-X(v'-2ik\gamma_0)\over
Y(v')}+2i\right)\nonumber\\
&&\times \left(\prod_{l=1\atop \neq k}B(v'-2li\gamma_0)\right) 
{(-1)^{N+m}D(v'-2ik\gamma_0)\over 2\sinh{1\over
2}(v'-v-2ik\gamma_0)}.
\label{dmaster}
\end{eqnarray}
These two operators may now be applied to the Bethe vector
(\ref{bv}). Then we  use (\ref{abit}) and (\ref{dbit}) 
to move $A(v'-2ik\gamma_0)$ and $D(v'-2ik\gamma_0)$ to the right
to act on $|0>,$ use the vanishing condition (\ref{vanish})
on all terms generated which contain $B(v'-2ik\gamma_0)$ and then 
add the two expressions together using the definition of the transfer matrix
in terms of $A(v)$ and $D(v)$ (\ref{tac})
to find
\begin{eqnarray}
&&T(v)B^{(N)}(v')\prod_{k=1}^nB(v_k)|0>=(-1)^{N+m}B^{(N)}(v')T(v)\prod_{k=1}^nB(v_k)|0>\nonumber\\
&&-B(v)\sum_{k=0}^{N-1}
\left\{\left({X(v'-2i(k-1)\gamma_0)-X(v'-2ik\gamma_0)\over
Y(v')}-2i\right)\right.\nonumber\\
&&\times\left(\prod_{l=1\atop \neq k}^N 
B(v'-2il\gamma_0)\right){(-1)^{N+m} {\tilde a}(v'-2ik\gamma_0)\over 2\sinh{1\over
2}(v'-v-2ik\gamma_0)} \nonumber\\
&&+\left({X(v'-2i(k+1)\gamma_0)-X(v'-2ik\gamma_0)\over
Y(v')}+2i\right)\nonumber\\
&&\left.\times \left(\prod_{l=1\atop \neq k}B(v'-2li\gamma_0)\right) 
{(-1)^{N+m}{\tilde d}(v'-2ik\gamma_0)\over 2\sinh{1\over
2}(v'-v-2ik\gamma_0)}\right\}\prod_{k=1}^nB(v_k)|0>
\label{master}
\end{eqnarray}
where
\begin{eqnarray}
{\tilde a}(v)&&=\sinh^L{1\over 2}(v-i\gamma_0)\prod_{k=1}^nf(v-v_k)\\
{\tilde d}(v)&&=\sinh^L{1\over 2}(v+i\gamma_0)\prod_{k=1}^nf(v_k-v).
\end{eqnarray}

In order for the defining relation for the operator $B^{(N)}(v)$ to hold
for $l=1$ the term proportional to $B(v)$ in (\ref{master}) 
must vanish. Thus we find  that $X(v)$ and 
$Y(v)$ must satisfy the following
functional equation for $k=0,\cdots ,N-1$
\begin{eqnarray}
&&\left({X(v-2i(k-1)\gamma_0)-X(v-2ik\gamma_0)\over
Y(v)}-2i\right){\tilde a}(v-2ik\gamma_0)\nonumber\\
&&+\left({X(v-2i(k+1)\gamma_0)-X(v-2ik\gamma_0)\over
Y(v)}+2i\right){\tilde d}(v-2ik\gamma_0)=0.
\label{funeq}
\end{eqnarray}
It is easily verified that this equation (and the periodicity
condition $X(v+2iN\gamma_0)=X(v)$ is satisfied if
\begin{equation}
{X(v)\over Y(v)}-{X(v-2i\gamma_0)\over Y(v)}
=-2i\left({N {\bar a}(v)\over \sum_{l=0}^{N-1}
{\bar a}(v-2il\gamma_0)}-1\right)
\label{xrecrel}
\end{equation}
with 
\begin{equation}
{\bar a}(v)=\sinh^L{1\over 2}(v-i\gamma_0)\prod_{k=1}^n
{1\over \sinh{1\over 2}(v_k-v)\sinh{1\over 2}(v_k-v+2i\gamma_0)}.
\end{equation}

The recursion relation (\ref{xrecrel}) only defines $X(v)$ up to an
additive function with  the periodicity $X_0(v-2i\gamma_0)=X_0(v).$ 
 But any such function $X_0(v)$ gives a vanishing contribution when
used in the expression for $B^{(N)}(v)$ (\ref{bcurrent})
by use of 
the relation which follows from
vanishing condition (\ref{vanish}) that
\begin{equation}
\sum_{k=0}^{N-1}
B_v(v-2ik\gamma_0)\prod_{l=0\atop \neq k}^{N-1}B(v-2li\gamma_0)=0.
\label{derived}
\end{equation}
Therefore 
we easily verify   by substitution that the only solution 
(\ref{xrecrel}) which gives a nonvanishing contribution to
$B^{(N)}(v)$ is
\begin{equation}
X(v)=2i\sum_{k=0}^{N-1}k{\bar a}(v-2ik\gamma_0).
\end{equation}
Therefore it also follows that
\begin{equation}
Y(v)=\sum_{k=0}^{N-1}{\bar a}(v-2ik\gamma_0)
\end{equation}
and thus (\ref{xres}) and (\ref{yres}) are proven. 

Finally we note that $Y(v)$ has a close connection with the eigenvalues
of the transfer matrix. To see this we note that from (\ref{teigen}) we find 
\begin{eqnarray}
&&(-1)^n{t(v)\over \prod_{k=1}^n\sinh{1\over
2}(v-v_k+2i\gamma_0)\sinh{1\over2}(v-v_k-2i\gamma_0)}\nonumber\\
&&={\sinh^L{1\over 2}(v-i\gamma_0)\over \prod_{k=1}^n
\sinh{1\over 2}(v-v_k-2i\gamma_0)\sinh{1\over 2}(v-v_k)}
+{\sinh^L{1\over 2}(v+i\gamma_0)\over \prod_{k=1}^n
\sinh{1\over 2}(v-v_k+2i\gamma_0)\sinh{1\over 2}(v-v_k)}.
\end{eqnarray}
Thus if we replace $v$ by $v-2il\gamma_0$ and sum on $l$ from $0$ to
$N-1$ we find that the two terms on the right hand side are equal by
periodicity and thus we obtain
\begin{equation}
(-1)^n{1\over 2}\sum_{L=0}^{N-1}{t(v-2il\gamma_0)\over \prod_{k=1}^n
\sinh{1\over 2}(v-v_k-2(l-1)i\gamma_0)\sinh{1\over
2}(v-v_k-2(l+1)i\gamma_0)}=Y(v)
\end{equation}
where we have used the expression for $Y(v)$ (\ref{yres}). 

\subsection{Multiple excitations}

It remains to demonstrate that the operator $B^{(N)}(v)$ which has been
shown to satisfy the defining equation (\ref{defining}) for $l=1$ 
in fact satisfies (\ref{defining}) for all integer $l>1.$ Consider
first $l=2.$ Then using (\ref{master}) to commute $T(v)$ to the right past
$B^{(N)}(v_1')$ we obtain 
\begin{eqnarray}
&&T(v)B^{(N)}(v_1')B^{(N)}(v_2')\prod_{k=1}^nB(v_k)|0>=\nonumber\\
&&+(-1)^{N+m}B^{(N)}(v_1')T(v)B^{(N)}(v_2')\prod_{k=1}^nB(v_k)|0>
\nonumber\\
&&-B(v)\sum_{k=0}^{N-1}
\left\{\left({X(v_1'-2i(k-1)\gamma_0)-X(v_1'-2ik\gamma_0)\over
Y(v_1')}-2i\right)\right.\nonumber\\
&&\times\left(\prod_{l=1\atop \neq k}^N 
B(v_1'-2il\gamma_0)\right){(-1)^{N+m} A(v_1'-2ik\gamma_0)\over 2\sinh{1\over
2}(v_1'-v-2ik\gamma_0)}B^{(N)}(v_2')\prod_{k=1}^nB(v_k)|0> \nonumber\\
&&+\left({X(v_1'-2i(k+1)\gamma_0)-X(v_1'-2ik\gamma_0)\over
Y(v_1')}+2i\right)\nonumber\\
&&\left.\times \left(\prod_{l=1\atop \neq k}B(v_1'-2li\gamma_0)\right) 
{(-1)^{N+m}D(v_1'-2ik\gamma_0)\over 2\sinh{1\over
2}(v_1'-v-2ik\gamma_0)}B^{(N)}(v_2')\prod_{k=1}^nB(v_k)|0>\right\}.
\label{master2}
\end{eqnarray}

In the first term on the right hand side of (\ref{master2}) we use
(\ref{master}) once again to (anti) commute $T(v)$ past $B^{(N)}(v'_2)$
without picking up any additional terms. In the remaining terms we
commute $A(v'_1)$ and $D(v'_1)$ to the right past $B^{(N)}(v'_2)$
using (\ref{amaster}) and (\ref{dmaster}). The vanishing 
condition (\ref{vanish})
shows that only the first terms in (\ref{amaster}) and (\ref{dmaster})
contribute and thus we find
\begin{eqnarray}
&&T(v)B^{(N)}(v_1')B^{(N)}(v_2')\prod_{k=1}^nB(v_k)|0>=\nonumber\\
&&+(-1)^{2(N+m)}B^{(N)}(v_1')B^{(N)}(v_2')T(v)\prod_{k=1}^nB(v_k)|0>
\nonumber\\
&&+(-1)^{N+m}B(v)B^{(N)}(v'_2)\sum_{k=0}^{N-1}
\left\{\left({X(v_1'-2i(k-1)\gamma_0)-X(v_1'-2ik\gamma_0)\over
Y(v_1')}-2i\right)\right.\nonumber\\
&&\times\left(\prod_{l=1\atop \neq k}^N 
B(v_1'-2il\gamma_0)\right){(-1)^{N+m} A(v_1'-2ik\gamma_0)\over 2\sinh{1\over
2}(v_1'-v-2ik\gamma_0)}\prod_{k=1}^nB(v_k)|0> \nonumber\\
&&+\left({X(v_1'-2i(k+1)\gamma_0)-X(v_1'-2ik\gamma_0)\over
Y(v_1')}+2i\right)\nonumber\\
&&\left.\times \left(\prod_{l=1\atop \neq k}B(v_1'-2li\gamma_0)\right) 
{(-1)^{N+m}D(v_1'-2ik\gamma_0)\over 2\sinh{1\over
2}(v_1'-v-2ik\gamma_0)}\prod_{k=1}^nB(v_k)|0>\right\}.
\label{master3}
\end{eqnarray}
The terms to the right of $B^{(N)}(v'_2)$ are now exactly the terms
shown to vanish in the previous section and thus (\ref{defining})
holds to $l=2.$ Iteration of this argument proves (\ref{defining}) for
arbitrary $l.$

\section{Properties of $X(v), Y(v)$ and the sum rules}

The operator $B^{(N)}(v)$ has the possibility of having poles at the poles
of $X(v)$ or the zeroes of $Y(v).$  

\subsection{The poles of $X(v)$}

The function $X(v)$ (\ref{xres}) has poles at
\begin{equation}
v=v_j+2ip\gamma_0~~~{\rm where}~~p=0,\cdots, N-1
\end{equation}
As $v\rightarrow v_j+2ip\gamma_0$ for $p \geq 1$ $X(v)$ behaves as
\begin{eqnarray}
X(v)\rightarrow&&{2ip\sinh^L{1\over 2}(v_j-i\gamma_0)\over
\sinh{1\over 2}(v_j+2ip\gamma_0-v)\prod_{k\neq j}^n\sinh{1\over
2}(v_k-v_j)\prod_k^n\sinh{1\over 2}(v_k-v_j+2i\gamma_0)}\nonumber\\
&&+{2i(p-1)\sinh^L{1\over 2}(v_j+i\gamma_0)\over
\sinh{1\over 2}(v_j+2ip\gamma_0-v)\prod_{k\neq j}^n\sinh{1\over
2}(v_k-v_j)\prod_k^n\sinh{1\over 2}(v_k-v_j-2i\gamma_0)}
\end{eqnarray}
which upon using the Bethe's equation (\ref{beq}) reduces to
\begin{equation}
X(v)\rightarrow{-2i\sinh^L{1\over 2}(v_j+i\gamma_0)\over
\sinh{1\over 2}(v_j+2il\gamma_0-v)\prod_{k\neq j}^n\sinh{1\over
2}(v_k-v_j)\prod_k^n\sinh{1\over 2}(v_k-v_j-2i\gamma_0)}.
 \end{equation}
 
As $v\rightarrow v_j$ we have
\begin{equation}
X(v)\rightarrow
{2i(N-1)\sinh^L(v_j+i\gamma_0)\over
\sinh{1\over 2}(v_j-v)\prod_{k\neq j}^n\sinh{1\over
2}(v_k-v_j)\prod_k^n\sinh{1\over 2}(v_k-v_j-2i\gamma_0)},
\end{equation}

Using these limiting forms in the expression for the operator
$B^{(N)}(v)$ and using (\ref{derived})
it follows that as $v\rightarrow v_j+2il\gamma_0$ for any $l$ that
\begin{eqnarray}
B^{(N)}(v)\rightarrow &&
{2iN\sinh^L{1\over 2 }(v_j+i\gamma_0)\over
\sinh{1\over 2}(v_j-v)\prod_{k\neq j}\sinh{1\over
2}(v_k-v_j)\prod_k\sinh{1\over 2}(v_k-v_j-2i\gamma_0)}\nonumber\\
&&\times B_v(v_j)\prod_{l=1}^{N-1}B(v_j-2li\gamma_0)
\end{eqnarray}
This operator vanishes when applied to the Bethe state because of the
vanishing condition (\ref{vanish}) and therefore
$B^{(N)}(v)$ does not have a pole
at $v=v_j+2ip\gamma_0$ even though $X(v)$ does.

\subsection{Properties of $Y(v)$}

The function $Y(v)$ defined by (\ref{yres}) also appears to have poles
at $v \rightarrow v_j+2ip\gamma_0.$ In this limit
\begin{eqnarray}
Y(v)
\rightarrow&&{\sinh^L{1\over 2}(v_j-i\gamma_0)\over
\sinh{1\over 2}(v_j+2ip\gamma_0-v)\prod_{k\neq j}^n\sinh{1\over
2}(v_k-v_j)\prod_k^n\sinh{1\over 2}(v_k-v_j+2i\gamma_0)}\nonumber\\
&&+{\sinh^L{1\over 2}(v_j+i\gamma_0)\over
\sinh{1\over 2}(v_j+2ip\gamma_0-v)\prod_{k\neq j}\sinh{1\over
2}(v_k-v_j)\prod_k^n\sinh{1\over 2}(v_k-v_j-2i\gamma_0)}.
\end{eqnarray}
However by use of the Bethe equation (\ref{beq}) we see that the
residue at this pole vanishes. Therefore $Y(v)$ in fact has no poles
and therefore must be a Laurent polynomial in $e^v.$ Moreover it
follows from the periodicity condition $Y(v+2i\gamma_0)=Y(v)$ that $Y(v)$
is in fact a Laurent polynomial in
\begin{equation}
z=e^{Nv}.
\end{equation}

To further study the properties of the Laurent polynomial we introduce
the positive and negative degrees of the polynomial defined by the
behavior of $Y(v)$ as $v\rightarrow \pm \infty $
as
\begin{equation}
Y(v)\sim C_{\pm} e^{\pm Nd_{\pm}v}~~{\rm as}~~v\rightarrow \pm \infty.
\end{equation}
We may thus define the Drinfeld polynomial as
\begin{equation}
P_{\Omega}(z)=e^{d_{-}Nv}Y(v).
\end{equation}
This is a polynomial in $z$ with degree
\begin{equation}
d=d_{+}+d_{-} 
\end{equation}

This limiting behavior of $Y(v)$ is obtained from (\ref{yres}).
There still are several cases to consider depending on 
the number of Bethe roots $v_k$ which take on the 
value of $\pm \infty.$ 

Consider first the case of no infinite $v_k.$ Then as $v\rightarrow
\pm \infty$ we find that 
\begin{eqnarray}
Y(v)\sim &&2^{-(L-2n)}\sum_{l=0}^{N-1}C_{\pm}(l)e^{|v|({L\over
2}-n-l)}\sum_{j=0}^{N-1}e^{-({L\over
2}-n-l)1\gamma_0(2j+1)}\nonumber\\
=&&2^{-(L-2n)}C_{\pm}(r)Ne^{|v|({L\over 2}-n-r)}
\label{yinf}
\end{eqnarray}
where
\begin{equation}
{L\over 2}-n-r\equiv 0~({\rm mod}~N)~~{\rm and}~~r=0,1,\cdots, N-1
\end{equation}
and the first few $C_{\pm}(j)$ are
\begin{eqnarray}
C_{\pm}(0)&&=\prod_{k=1}^ne^{\pm v_k}\label{sum0}\\
C_{\pm}(1)&&=\left(-L+(e^{i\gamma_0}+e^{-i\gamma_0})\sum_{k=1}^n e^{\pm
v_k}\right)\prod_{k=1}^n e^{\pm v_k}\label{sum1}\\
C_{\pm}(2)&&=\left({1\over
2}L(L-1)-L(e^{i\gamma_0}+e^{-i\gamma_0})\sum_{k=1}^ne^{\pm
v_k}\right.\nonumber\\
&&\left.+{1\over 2}(e^{2i\gamma_0}+e^{-2i\gamma_0})\sum_{k=1}^n e^{\pm2v_k}
+{1\over 2}(e^{i\gamma_0}+e^{-i\gamma_0})^2(\sum_{k=1}^ne^{\pm v_k})^2\right)
\prod_{k=1}^n e^{\pm v_k}
\label{sum2}
\end{eqnarray}
Therefore for the case of no infinite roots $v_k$ 
we have
\begin{equation}
d_{+}=d_{-}=({L\over 2}-n-r)/N=[S^z/N]
\end{equation}
where in the last line we have used (\ref{spin}) and
$[x]$ denotes the greatest integer in $x$ 
Thus the degree of the Drinfeld polynomial 
for highest weight states with no infinite roots is
\begin{equation}
d=2[S^z/N].
\end{equation}

When $S^z\equiv 0~({\rm mod}~N)$ all evidence in
refs. \cite{paper1}-\cite{paper3} is that infinite roots never
occur. However for all other cases there are indeed infinite roots and
the evidence of refs. \cite{paper1}-\cite{paper3} is that there are
two cases to distinguish

(A) The pair $v=\infty$ and $-\infty$ occurs $p$ times.

(B) The single root $v_k=\infty$ or $-\infty$ occurs $s$ times.

For case (A) we let $p$ pairs of roots go to $\pm \infty$ in
(\ref{xres}) and (\ref{yres}) and define
\begin{equation}
\lim {X(v)\over Y(v)}={X_p(v)\over Y_p(v)}
\end{equation}
where $X_{p}(v)$ and $Y_p(v)$ are obtained from (\ref{xres}) and
(\ref{yres}) by merely omitting the infinite roots (and thus replacing
$n$ by $n-2p).$

The previous argument generalizes by
replacing $n$ by $n-2p$ and thus we find
\begin{equation}
Y_p(v)\sim2^{-(l-2n+4p)}C_{\pm}(r)Ne^{|v|({L\over 2}-n-r+2p)}
\end{equation}
where
\begin{equation}
{L\over 2}-n-r+2p\equiv 0~({\rm mod}~N).
\label{helpdeg}
\end{equation}
From the data of table 10 of ref. \cite{paper2} for $N=3$ we find that
for $S^z\equiv 1~({\rm mod}~3)$ that $p=2$ and for $S^z\equiv 2~({\rm
mod}~3)$ that $p=1.$ This gives $r=p$ for $N=3.$ If this holds
generally for all $N$ then we find from (\ref{helpdeg}) that the degrees of
$Y_p(v)$ 
are
\begin{equation}
d_{+}=d_{-}=[S^z/N]+1
\end{equation}
and thus the degree of the Drinfeld polynomial is
\begin{equation}
d=2[S^z/N]+2
\end{equation}

For case (B) we let $s$ roots $v_k\rightarrow \infty$ or $-\infty$
in (\ref{xres}) and (\ref{yres}) to find
\begin{eqnarray}
X(v)_{\pm s}&&=2i\sum_{l=0}^{N-1}
\frac{l e^{\mp s(2l+1)i\gamma_0}
\sinh^L{1\over 2}(v-(2l+1)i\gamma_0)}
{\prod_{k=1}^{n-s}\sinh{1\over 2}(v-v_k-2il\gamma_0)\sinh{1\over
2}(v-v_k-2i(l+1)\gamma_0)}
\label{sxres}
\end{eqnarray}
and
\begin{eqnarray}
Y(v)_{\pm s}=&&\sum_{l=0}^{N-1}\frac{e^{\mp s(2l+1)i\gamma_0}
\sinh^L{1\over 2}(v-i\gamma_0(2l+1))}
{\prod_{k=1}^{n-s}\sinh{1\over 2}(v_k-v+2il\gamma_0)
\sinh{1\over 2}(v_k-v+2i(l+1)\gamma_0)}
\label{syres}
\end{eqnarray}
Now the limiting behaviors as $v\rightarrow \pm \infty$ are no longer
the same and we find from (\ref{syres}) that as $v\rightarrow \infty$
\begin{equation}
Y_{\pm s}(v)\sim 2^{-(L-2n-2s)}C_{\pm}(r_{+})Ne^{|v|({L\over 2}-n-r_{+}+s)}
\end{equation}
where
\begin{equation}
{L\over 2}-n-r_{+}+s\pm s\equiv 0~({\rm mod}~N).
\end{equation}
and similarly as $v\rightarrow -\infty$ 
\begin{equation}
Y_{\pm s}(v)\sim 2^{-(L-2n-2s)}C_{\pm}(r_{-})Ne^{|v|({L\over 2}-n-r_{-}+s)}
\end{equation}
with
\begin{equation}
{L\over 2}-n-r_{-}+s\mp s\equiv 0~({\rm mod}~N).
\end{equation}
In ref. \cite{paper2} we found that for $S^z=1~({\rm mod}~ 3)$ that
$s=2$ and if $S^z=2~({\rm mod}~ 3)$ that $s=1.$
From this we find  for $Y_{+s}$ if $S^z\equiv
1~({\rm mod}~3)$  that $r_{+}=2,~r_{-}=1$ and if
$S^z\equiv 2~({\rm mod}~3)$  that $r_{+}=1,~r_{-}=2.$
Thus we find for this (and for all other $N$ with the same relation of
$r_{\pm}$ to $S^z$ that
\begin{equation}
d_{+}=[S^z/N]+1,~~d_{-}=[S^z/N]
\end{equation}
and thus the degree of the Drinfeld polynomial is
\begin{equation}
d=2[S^z/N]+1.
\end{equation}
\subsection{Sum rules for Bethe roots}

In general the Laurent polynomial $Y(v)$ has $d_{+}+d_{-}\geq
1$. However if the highest weight state has no infinite roots and if
\begin{equation}
0\leq S^z\leq N-1
\label{szres}
\end{equation}
then we see from (\ref{yinf}) that $Y(v)$ is a constant which must equal the
limiting values obtained from $v\rightarrow \pm \infty.$  Thus it
follows that when (\ref{szres}) holds we have $N$ distinct sum rules
\begin{equation}
C_{+}(S^z)=C_{-}(S^z).
\end{equation} 
For example we find from (\ref{sum0})-(\ref{sum2}) that
\begin{eqnarray}
S^z=0:&&~~\sum_{k=0}^{L/2}v_k=0,~{\rm or}~\pi i\label{sumrule0a}\\
S^z=1:&&~~
\{L-(e^{i\gamma_0}+e^{-i\gamma_0})\sum_{k=1}^{{L\over2}-1}e^{v_k}\}
            \prod_{k=1}^{{L\over2}-1}e^{v_k}
          =\{L-(e^{i\gamma_0}+e^{-i\gamma_0})\sum_{k=1}^{{L\over 2}-1}
e^{-v_k}\}
            \prod_{k=1}^{{L\over2}-1}e^{-v_k}\label{sumrule1a}\\
S^z=2:&&
\left({1\over
2}L(L-1)-L(e^{i\gamma_0}+e^{-i\gamma_0})\sum_{k=1}^ne^{
v_k}\right.\nonumber\\
&&\left.+{1\over 2}(e^{2i\gamma_0}+e^{-2i\gamma_0})\sum_{k=1}^n e^{2v_k}
+{1\over 2}(e^{i\gamma_0}+e^{-i\gamma_0})^2
(\sum_{k=1}^ne^{v_k})^2\right)\prod_{k=1}^n e^{v_k}\nonumber\\
&&=\left({1\over
2}L(L-1)-L(e^{i\gamma_0}+e^{-i\gamma_0})\sum_{k=1}^ne^{-
v_k}\right.\nonumber\\
&&\left.+{1\over 2}(e^{2i\gamma_0}+e^{-2i\gamma_0})\sum_{k=1}^n e^{-2v_k}
+{1\over 2}(e^{i\gamma_0}+e^{-i\gamma_0})^2
(\sum_{k=1}^ne^{-v_k})^2\right)\prod_{k=1}^n e^{- v_k}
\end{eqnarray}

These sum rules have been derived under the assumption that $\gamma_0$
satisfies the root of unity (rationality) condition
(\ref{root}). However, for fixed $S^z,$ these sum rules hold for all
$N$ such that $S^z\leq N-1$ and will therefore hold  for all $\gamma$
such that $\gamma/\pi$ is irrational. Moreover the sum rule will also
hold at all rational values of $\gamma/\pi$ for the Bethe's roots
which are obtained by continuity from the irrational values of
$\gamma/\pi$ provided that all the $v_k$ are finite. Thus we have
proven the general sum rules (\ref{sumrule0})-(\ref{sumrule2})
given in the introduction.

It is worth noting that these sum rules follow merely from the
properties of the function $Y(v)$ as defined in terms of the Bethe
roots by (\ref{yres}). The relation of $Y(v)$ to the  operator
$B^{(N)}(v)$ has not been used in this derivation and in principle
these sum rules, which to the authors knowledge are new, could have
been derived without the knowledge of the $sl_2$ loop algebra symmetry
which forms the starting point for this present paper.

\section{Representation Theory}

The mathematical theory of finite dimensional representations of affine Lie
algebras \cite{char}-\cite{cp} and the demonstration \cite{paper1} of the
$sl_2$ loop algebra symmetry in the six vertex model at roots of unity
in the sector $S^z\equiv 0~({\rm mod}~N)$ makes no use of the
algebraic Bethe's ansatz. On the other hand the computation of the
$B^{(N)}(v)$ in this paper makes no use of representation
theory. Nevertheless the conjecture that 
\begin{equation}
B^{(N)}(v)=N(z)E^{-}(z),
\end{equation}
where $N(z)$ is a scalar with no poles implies that
the two methods are part of the same subject and
that by combining them we are able to obtain 
results which have so far been inaccessible to either method in
isolation. 

One such result is the explicit expression for the Drinfeld
polynomial 
\begin{equation}
P_{\Omega}(z)=e^{d_{-}Nv}Y(v).
\end{equation}
 
Furthermore if the roots of $Y(v)$ are distinct
the current $E^{-}(z)$ will be given solely in terms of the residue at the poles
$z=a_j$ where $P_{\Omega}(a_j)=0.$ Therefore is must follow that 
\begin{equation}
{B^{(N)}(v)\over N(z)}|\Omega>=
\sum_{j=1}^{d_{\Omega}}{zE^{(N)}_{\Omega}(j)\over P'_{\Omega}(a_j)(z-a_j)}
|\Omega>
\label{ecurrent}
\end{equation}
When the roots $P_{\Omega}$ are distinct representation theory says
that only spin 1/2 representations occur in the  degenerate
multiplets and therefore
\begin{equation}
E^{(N)2}_{\Omega}(j)|\Omega>=0.
\label{spinhalf}
\end{equation}
From this we see that if we denote the spin in the multiplet by
\begin{equation}
S^z=S^z_{\rm max}-Nl ~~{\rm with}~~0\leq l \leq d_{\Omega}
\end{equation}
that the
multiplicity is given by the binomial coefficient given in the
introduction (\ref{binomial})
\begin{equation}
{\rm multiplicity}={d_{\Omega}\atopwithdelims() l}.
\end{equation}
This is in complete agreement with ref. \cite{paper1} for all
sectors not only for $S^z\equiv 0~({\rm mod}~N)$ where the symmetry
algebra was proven but in all other sectors where a projection was
needed in order to obtain the algebra. Thus the mechanism of the infinite
roots which automatically appears in the algebraic Bethe ansatz
makes explicit the projection mechanism only effected on the computer
when using the Chevalley basis for the loop algebra generators \cite{paper1}.
We also note from (\ref{ecurrent}) that the
operator $B^{(N)}(v)$ acting on $|\Omega>$ 
will create a vector space of dimension
$d_{\Omega}.$ For the case $N=3$ this dimension has been computed from
the coordinate space Bethe's ansatz by Braak and Andrei \cite{ba}.

\section{Conclusions}

The results of the previous section are obtained by using 
the representation theory of affine Lie algebras \cite{char}-\cite{cp} 
to provide
existence theorems for algebraic Bethe's ansatz computations. It is
clearly desirable  to avoid representation theory altogether and to
explicitly compute the normalizing constant $N(z)$ and the residue
operators $E_{\Omega}^{(N)}(j)$ explicitly from $B^{(N)}(v).$
Conversely it is desirable to compute the expression (\ref{yres}) for the
Drinfeld polynomial directly from representation theory without the use
of the algebraic Bethe's ansatz. 
These can perhaps be viewed as mathematical problems.

But from the physical point of view perhaps the most interesting
question is to inquire into the physical meaning and significance of
the Drinfeld polynomial (\ref{yres}). This polynomial has been seen previously
in our study  of the limiting form of Bethe's ansatz as 
$\gamma\rightarrow \gamma_0$ (see the function $K(\alpha_j)$
of eqn. 2.18 of ref. \cite{paper3}).
In addition  very closely related expressions have been seen in the 
computation  of the free
energy of the superintegrable chiral Potts model (see eqn. 4.5 of
ref. \cite{amp}) and in the study of the RSOS models (see eqn. 3.29 of
ref.\cite{br}).
The evaluation parameters, which are the roots of the
Drinfeld polynomial, must have a further physical significance which
is yet to be discovered.

Finally it needs to be noted that everything in this paper can 
be extended from the spin 1/2 XXZ model to any model based
on a quantum group just as the loop symmetry was extended in ref.\cite{km}. 

\vspace{.2in}
\newpage 

\centerline{{\large \bf Acknowledgments}}

\vspace{.1in}

We are very pleased to acknowledge fruitful discussions with
R.J. Baxter, E. Date, T. Deguchi, M. Jimbo, M Kashiwara, C. Korff, A. Kuniba, 
T. Miwa, and K. Okado. We also thank M. Kashiwara and T. Miwa for
their invitation to ``MathPhys Odyssey 2001 Integrable Models and
Beyond'' where this work began. 
One of us (BMM) is pleased to 
thank M. Kashiwara and the
Research Institute of Mathematical Sciences of Kyoto University for their
hospitality. This work was supported in part by the National Science
Foundation under Grant DMR-0073058.

\end{document}